\documentclass[twocolumn]{aastex631}

\usepackage{amsmath,amssymb}

\usepackage{overpic}
\usepackage{tikz}

\graphicspath{{figs/}}

\accepted{April 27, 2025}

\newif\ifref
\reffalse
\definecolor{darkred}{rgb}{0.75, 0, 0}
\newcommand{\mb}[1]{\ifref\textcolor{darkred}{#1}\else #1\fi}

\newif\ifreff
\reffalse
\definecolor{darkred}{rgb}{0.75, 0, 0}
\newcommand{\mbb}[1]{\ifreff\textcolor{darkred}{#1}\else #1\fi}

\submitjournal{the Astrophysical Journal}

\begin{document}

\title{The Sun's Dark Core: Helioseismic and neutrino flux constraints on a compact solar center}
\shorttitle{The Sun's Dark Core}
\shortauthors{Bellinger \& Caplan}

\correspondingauthor{Earl P.\ Bellinger}
\email{earl.bellinger@yale.edu}

\author[0000-0003-4456-4863]{Earl P.\ Bellinger} 
\affil{Department of Astronomy, Yale University, CT, USA} 

\author[0000-0000-0000-0000]{Matt E.\ Caplan} 
\affiliation{Department of Physics, Illinois State University, IL, USA}

\begin{abstract}
As dark matter appears to comprise most of the Galactic mass, some of it may accumulate in the cores of stars, thereby making the Sun a laboratory for constraining various dark matter theories. 
We consider the effects on the solar structure arising from a general class of macroscopic dark matter candidates that include strange quark matter, compact dark objects, and others. 
We calibrate standard solar evolution models (i.e., models that reproduce the mass, luminosity, radius, and metallicity of the Sun at its present age) with variable compact dark core masses ranging from $10^{-8}$ to $10^{-2}~\rm{M}_\odot$ and assess their properties.   

We find that the weakest constraints come from solar neutrino flux measurements, which only rule out the most massive dark core comprising at least $\sim 1\%$ of the total solar mass. 
The Sun's acoustic oscillations impose stronger constraints, probing masses down to $\sim 10^{-5}~\rm{M}_\odot$. 
We find that a model with a $10^{-3}~\rm{M}_\odot$ dark core appears to improve the agreement with helioseismic observations. 
We nevertheless caution against interpreting this as evidence for dark matter in the solar interior, and suggest plausible effects that the dark core may instead be emulating. 
Finally, we show that future measurements of solar $g$~modes may constrain dark core masses down to $10^{-7}~\rm{M}_\odot$. 
\end{abstract}%
%
\section{Introduction} \label{sec:intro}
The development of a standard model of the Sun is among the greatest successes of 20th century physics. 
The model evolved iteratively over many decades, and required major advances in nuclear and particle astrophysics in order to reproduce the basic properties of the Sun. 
The model's greatest successes were its prediction of the neutrino production and internal sound speed profile of the Sun. The latter is within 1\% of the values inferred from helioseismology, demonstrating that the internal structure of the solar model is in broad qualitative agreement with the structure of the actual Sun. 
Nevertheless, the standard solar model remains to this day significantly discrepant with helioseismic observations, which has motivated many explorations into possible ways to potentially improve the model. 

The standard solar model has a fascinating history involving many of the most pivotal figures in astronomy. 
Early models of the Sun assumed the same composition as the Earth and relied on gravitational contraction as the source of their energy \citep{kelvin1862, ARNY1990211}. 
These models started as giants and ended their lives on the main sequence \citep{ritter}. 
The age of such a solar model however was incompatible with meteoric dating, and corresponding stellar models were incompatible with observations of pulsating stars \citep{1920Natur.106...14E}. 

Stellar astrophysics underwent a scientific revolution in the sense of \citet{Kuhn1962} following the argument of \citet{1920Natur.106...14E} that subatomic energy must supply a star's energy. 
In combination with the discovery of \citet{1925PhDT.........1P} that stars are primarily made of hydrogen, \citet{1938PhRv...53..907G} constructed models with the entire star undergoing nuclear fusion. 
However, these models had serious problems, requiring decades of revision that continue to today. 

The first generation of stellar evolution models powered by nuclear fusion failed to turn into giants, giving no clear route to explain their origin. 
\citet{1938PTarO..30C...1O} discovered that a layered stellar structure remedied this problem: stars may be first powered by core fusion on the main sequence, and subsequently by fusion in a shell surrounding the inert core, thereby enabling stars to climb the red giant branch after the long main sequence. 
Solar models thereafter only fused hydrogen in their central regions. 
\citet{1946ApJ...104..203S} published one of the first numerical solar models, which was powered by CNO fusion in a convective core with a radiative envelope --- the opposite of the situation we recognize for the present Sun. One decade later, \citet{1957ApJ...125..233S} revised his model to have the Sun gain its energy from the proton-proton chain. 
Along with other improvements such as in our understanding of the equation of state and opacity of solar plasma, these developments culminated in the standard model of the Sun: a star born of a uniform composition of mostly hydrogen, powered by core fusion, and transporting its energy via radiation in the interior and convection in its envelope \citep{1972ARA&A..10...25B}. 

Soon following the development of such a standard solar model, new observations began challenging it. 
In the early 1960s, the global oscillations of the Sun were discovered \citep{1962apj...135..474l}, enabling detailed insight into its internal structure \citep{1976Natur.259...89C}. 
In the following decade, \citet{1976Sci...191..264B} measured neutrinos from solar nuclear reactions, leading to the \mb{2002} Nobel Prize in Physics. 
These measurements gave rise to two major problems in astrophysics. 
The first is the now-solved ``solar neutrino problem'': neutrino counts were much lower than predicted by stellar evolution theory. 
This was later remedied by the discovery of neutrino flavor oscillations \citep{2002PhRvL..89a1301A}, leading to another Nobel Prize. 
Still unsolved is the ``solar abundance problem'' or more broadly, the ``solar modeling problem'': the standard solar model remains highly statistically discrepant with measurements of the solar oscillations. 
Although the internal sound speed profile only differs with helioseismic measurements by about 0.5\%, the measurements of the solar sound speed profile are precise to a level of 0.005\%, yielding a model that is in high tension with the measurements, despite their broad qualitative agreement. 
The consensus today is that the current standard solar model does not have exactly the right internal structure. 
This problem has only worsened with more recent determinations of the solar abundances \citep[e.g.,][]{2021LRSP...18....2C, 2023A&A...669L...9B}. 
Helioseismic and neutrino flux observations therefore appear to confirm the general picture of solar evolution, but nevertheless show that that the standard model of the Sun still has serious issues. 

As this problem remains outstanding for more than four decades, a large literature has developed on various attempts to improve the structure of the solar model. Here we will review but a few attempts; see \citealt{2016LRSP...13....2B} and \citealt{2021LRSP...18....2C} for comprehensive reviews. 
\citet{1993apj...403l..75c} discovered that heavy element settling in the solar interior reduces the discrepancy, but does not eliminate it. 
\citet{1994MNRAS.267..209B} assessed whether mixing beyond the base of the solar convection zone would improve the model, but found that the amount of overshooting must be very small. 
\citet{2009A&A...494..205C} considered whether changes to opacities could improve agreement with observations, particularly around the base of the solar convection zone. 
\citet{2024ApJ...968...56F} recently computed extensive new tables of opacities, and concluded that they do not fix the solar model. 
\citet{2022MNRAS.517.5281B} assessed whether uncertainties in nuclear reaction rates could be the culprit, and found that while helioseismic data in principle constrain nuclear reaction rates better than laboratory measurements, adjustments to these rates also do not solve the problem. 
\citet{2019ApJ...881..103Z} found that the discrepancy is reduced if the Sun has a metal-rich core compared to its envelope, and \citet{2021A&A...655A..51K} and \citet{2022A&A...667L...2K} have argued that this could arise from ingestion of a metal-poor protosolar disk during the later part of its formation period. 

With this problem still unsolved, it motivates additional approaches to try and remedy the solar structure, as well as explorations into what physics can be constrained using present and anticipated future helioseismic and neutrino flux measurements. 
Here we focus on one such avenue: dark matter. 

\subsection{Particle Dark Matter Solar Models}
The problem of dark matter is among the most notable unsolved problems in astrophysics. 
Numerous observations indicate that our Universe appears to be permeated by enormous quantities of so-far invisible matter that interacts gravitationally with normal matter. 
Furthermore, modern cosmological theories require dark matter in order to successfully predict the large-scale structure of the Universe as well as the formation and structure of the Milky Way \citep[e.g.,][]{2023ApJS..265...44W}. 

In the $\Lambda$CDM paradigm, the Milky Way initially formed through baryons falling into the dark matter halo that comprises $\sim$84\% of the total Galactic mass \citep[e.g.,][]{2010gfe..book.....M}. 
Consequently, stellar systems within the Milky Way, including the solar system, receive a constant flux of this dark matter, some of which swarms, orbits, and potentially accumulates in the center of stars \citep{1987ApJ...321..571G, 2009ApJ...705..135C, 2014JCAP...05..049C, 2015JCAP...04..042C, 2024arXiv240508113D}. 
This has given rise to numerous ventures to assess whether various dark matter candidates can be detected through their effects on the properties of the Sun. 
These include ground-based detectors, such as the CERN Axion Solar Telescope \citep{2005PhRvL..94l1301Z, 2017NatPh..13..584A}, as well as investigations into the changes to the solar structure, pulsations, and neutrino flux that arise depending on the assumed microphysics of dark matter \citep[e.g.,][]{2022FrASS...9.8502A}.  
As the origin and nature of dark matter still remain unknown, many studies focus on the specific effects of particular candidates, which have included hidden photons, weakly-interacting massive particles (WIMPs), axions, asymmetric dark matter (ADM), supersymmetric dark matter, light dipole dark matter, and fuzzy dark matter (for a recent review of dark matter, see \citealt{2024arXiv240601705C}). 

An early investigation by \citet{1990ApJ...353..698C} studied the effects of WIMPs on solar models and found that they significantly reduce the frequency differences between radial and quadrupole modes. 
However, \citet{1991ApJ...378..315K} found that the effect of WIMPs on the sound speed near the solar center did not improve alignment with helioseismic data.
\citet{2002MNRAS.337.1179L, 2002MNRAS.331..361L} explored the impact of WIMP accretion and annihilation on solar evolution, noting that WIMPs with masses less than 30 GeV conflict with seismic data due to their impact on the solar core's temperature, density, and composition. The effect is negligible for WIMPs with higher masses due to their decreased abundance. 
\citet{2002PhRvD..66e3005B} found that WIMP interactions alter the squared isothermal sound speed by up to 0.4\% and reduce the $^8$B neutrino flux by 4.5\%, but reconfirmed that current helioseismology and neutrino data do not strongly constrain WIMP properties.
\citet{2012ASPC..462..537H} and \citet{2013JCAP...08..034R} studied the effects of light WIMPs on solar structure, and indicated that WIMPs below 10 GeV are ruled out by constraints on core sound speed and low-degree frequency spacings.

\citet{2010PhRvL.105a1301F} discussed ADM particles and suggested they could solve the solar composition problem, with a predicted small decrease in low-energy neutrino fluxes.
\citet{2010PhRvD..82h3509T} investigated these further and concluded that detectable effects on neutrino fluxes only occur in models with very small or vanishing self-annihilation cross sections.
\citet{2014ApJ...780L..15L} suggested that the presence of ADM with a spin-independent scattering cross section could induce changes in solar neutrino fluxes and helioseismology data, particularly if dark matter particles have masses below 15 GeV. 

\citet{2015JCAP...10..015V} combined helioseismology and solar neutrino observations to place upper limits on the properties of non-standard weakly interacting particles, improving constraints on axion-photon coupling and hidden photon parameters. 
\citet{2010PhRvD..82j3503C} and \citet{2012ApJ...757..130L} studied non-annihilating dark matter, and found that such particles filling the solar core could cause changes in $g$-mode period spacings of up to 20\%. 
\citet{2017JCAP...05..007G} examined self-annihilating DM scattering off electrons versus nucleons in the Sun, suggesting that electron interactions could dominate under certain conditions, impacting neutrino production rates. 
\citet{2024MNRAS.527L..14S} proposed that ultra-light bosons, such as axions or fuzzy dark matter, could resonantly excite solar oscillations, though found that the predicted oscillations are likely undetectable. 

These studies highlight the potential for the Sun to play an important role in our understanding of dark matter. There are however yet more dark matter candidates whose effects on the Sun have not yet been considered, motivating the present work. 

\subsection{Macroscopic Dark Matter}
In this work, we do not consider a specific dark matter candidate, and remain agnostic about its fundamental nature. 
Rather, we are concerned with the general class of dark matter candidates near or above nuclear densities that are macroscopically massive. 
Such candidates are distinct from the loose collection of particles described above. 
Consequently, we neglect particle-level properties such as mass, Standard Model interaction cross-sections, or rates of self-interaction.

While macroscopic dark matter could consist of new particles, there are also pathways from the Standard Model to produce them, such as clusters of strange quark matter (\citealt{1984PhRvD..30..272W}; for a recent review, see \citealt{diclemente2024strange}). 
Similarly, a wide range of new particles in a dark sector could form composite objects. 
For example, beyond-Standard Model confining gauge theories could form `dark quark matter' with nuclear densities and almost arbitrary macroscopic masses. Dark sector particles that experience some weak dissipation via a dark force, but do not self-annihilate, could in principle form cuspy halos that contract to become effectively point-like.

Tests of this broad class of potential dark matter (with macroscopic masses and high compactness) have been widely considered and are known by a variety of names in the literature, including compact dark objects (CDOs), compact ultra-dense objects (CUDOs), and macros \citep{rafelski2013compact,jacobs2015macro,Horowitz_2020}. 
These candidates therefore motivate us to consider the effects of a non-luminous central point mass in the solar center that only interacts gravitationally with the remainder of the Sun. 
It is worth emphasizing that primordial black holes (PBHs) are not considered in the present work, because the PBH mass would be time-dependent as it accretes the solar plasma, and this accretion process may additionally provide some point-like luminosity in the core \citep{1995MNRAS.277...25M, bellinger2023solar, caplan2024there, 2024arXiv240617052S}. 

\subsection{Solar Models with Macroscopic Dark Matter}
We focus on models of the Sun that have a massive compact core, which we achieved by inserting a point mass of various masses at the center of the solar model. 
Mass distributions with central point-masses have been previously considered for modeling galactic nuclei with supermassive black holes in their centers \citep{huntley1975distribution}. 
These so-called `loaded polytropes' can equivalently be used for stars with an appropriately chosen polytropic equation of state. 
Most generally, the equation for hydrostatic equilibrium with an additional term for a point mass at the origin is 
\begin{equation}
    \frac{\textrm{d}P}{\textrm{d}r} 
    = 
    -
    \frac{G M_0 \rho(r)}{r^2} 
    -
    \frac{G M(r) \rho(r)}{r^2}
\end{equation}
using pressure $P$ and density $\rho$, with $M_0$ the central mass and $M(r)$ the enclosed stellar mass (i.e., not including the dark core) at radius $r$.  
Solutions to the Lane-Emden equation (i.e., models that satisfy hydrostatic equilibrium with a polytropic equation of state) have density cusps around the central point mass. 
The cusp contains a mass comparable to the central point mass, such that when $M(r) \lesssim M_0$ one finds a cusp, and when $M(r) \gtrsim M_0$ the mass of the central mass is sufficiently diluted and the density is the roughly constant core density as one expects for an $n=3$ polytrope \citep{huntley1975distribution}.

\begin{figure*}
    \centering
    \includegraphics[trim={0 0 67 0},clip,height=9cm]{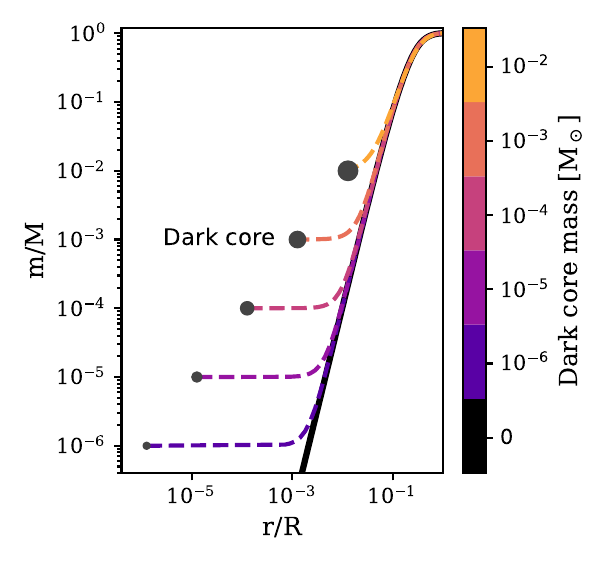}
    \includegraphics[height=9cm]{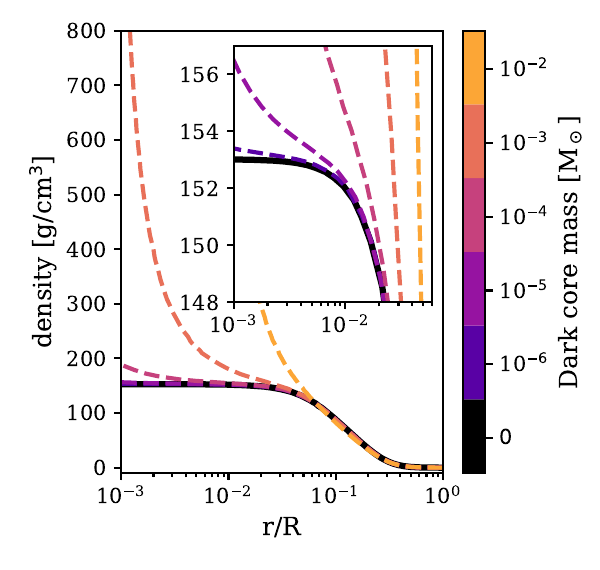}%
    \caption{\textbf{Internal structures of calibrated dark core solar evolution models.} These models have the mass, radius, luminosity, metallicity, and age of the Sun, as well as a variable amount of macroscopic dark matter represented by a point mass at their center. 
    \textsc{Left Panel}: Mass distribution of dark core models. 
    \textsc{Right Panel}: Density profiles of dark core models. The inset shows a zoom into the core, showing also the normal solar model with its central density of $\sim 153$~g/cm$^3$.
    \label{fig:mr}}
\end{figure*}

We have in recent work developed the stellar evolution code \textsc{Mesa} to simulate stars with central PBHs, which we adapt for use in this work. 
\cite{bellinger2023solar}, \cite{caplan2024there}, and \cite{2024arXiv240617052S} set the mass at the inner boundary to be the primordial black hole mass, in this work taken more generally to be the dark core mass. 
We follow those previous works and truncate the integration at the Bondi radius, $R_{\rm B} = 2 G M_0/c_s^2$, where $c_s$ is the speed of sound \citep{bondi1952spherically,bellinger2023solar}. 
\mb{The Bondi radius, when expressed in solar radii, has approximately the same value as the dark core mass expressed in solar masses.} 
Thus for the largest $10^{-2}~\rm{M}_\odot$ that we consider in this work, the Bondi radius corresponds to 1\% of the solar radius. 
Figure~\ref{fig:mr} shows the density structure of our dark core solar models. 

Unlike previous works, we do not prescribe accretion or luminosity to the central point mass, and we consider only the effects on evolutionary models of the present Sun. \mbb{Any constraints from these models are therefore restricted to dark matter candidates that would be non-accreting when embedded in baryonic matter. This allows us to probe dark matter models that are truly dark, such as beyond Standard Model particles that are non-interacting electromagnetically. Additionally, it also models hypothetical Standard Model arrangements of particles that could be stable and in hydrostatic equilibrium, such as electrically neutral or positively charged strangelets \citep{jaffe2000review}. However, we emphasize that this is a strength, as accretion rates and luminosities have large uncertainties such that non-accreting dark cores can be constrained much more reliably with existing methods.\footnote{\mbb{Multiple observations independently rule out stable electrically negative strangelets that would catastrophically accrete and convert baryonic matter, including cosmic ray data, high energy collider experiments, and the continued existence of the Moon \citep{jaffe2000review}.}}}

\mb{This model of non-accreting dark matter considers dark matter and baryonic matter that coexist in the core and are effectively non-interacting beyond gravity. 
This means our point-like mass at the core probes any dark matter that is relativistically compact with some small spatial extent (i.e., less than the Bondi radius). 
This produces large but not infinite pressures and densities in the core. 
The baryonic matter reaches high densities in response, providing hydrostatic equilibrium. 
The cusp of core baryonic matter would reach central densities with large electron degeneracy pressure, possibly even neutron star densities in a very small spatial extent. 
For example, \citet{Horowitz_2020b} considered similar compact dark matter cores in white dwarfs (see their Fig.~3).}

In the following sections, we detail the construction of the dark core models, their properties, and their comparison with solar observations. 
We find that the main effect of the presence of macroscopic dark matter in the solar core is a significantly stronger gravitational potential in the innermost regions of the star. 
This induces rapid settling, causing the models to have a heavy metal core. 
Additionally, pp fusion moves off-center, and CNO reactions increase. 

We assess neutrino fluxes and measurements of the Sun's radial structure as inferred through $p$-mode helioseismology. 
We encounter the unexpected result that a model with a $10^{-3}~\rm{M}_\odot$ dark core is in better agreement with helioseismic observations than the standard model. 
We argue that this arises due to the heavy metal core, and rather than being evidence for dark matter, we suggest that the dark matter prescription may be emulating some effects of star formation that are typically neglected in solar evolution simulations. 
We nevertheless show that the dark core model makes additional predictions that would clearly distinguish it if they would be observed; namely, the model has a large spectrum of mixed oscillation modes, and additionally has a significantly reduced $g$-mode period spacing. 
Therefore, future measurements may be able to confirm or further constrain the mass of the Sun's dark core.

\section{Solar calibration} \label{sec:calibration}
We compute the evolution of solar models using the open-source software instrument Modules for Experiments in Stellar Astrophysics \citep[\textsc{Mesa} version r24.03.1,][]{Paxton2011, Paxton2013, Paxton2015, Paxton2018, Paxton2019, Jermyn2023}. We use the FreeEOS equation of state, opacities from OPAL \citep{1996ApJ...464..943I}, nuclear reaction rates from JINA REACLIB \citep{Cyburt2010} and NACRE \citep{Angulo1999}, and  screening via the prescription of \citet{Chugunov2007}. We include the effects of atomic diffusion and gravitational settling in the evolution by solving the \citet{1969fecg.book.....B} equations. 
We neglect rotation, overshoot, magnetic fields, and mass loss. 

Normally in stellar evolution calculations, the stellar structure equations are solved using the inner boundary conditions 
\begin{equation}
    m_0=0, \qquad r_0=0, \qquad l_0=0
\end{equation}
where $m_0$ is the enclosed mass at the center, $r_0$ the distance from the center, and $l_0$ the central luminosity \citep[e.g.,][]{2013sse..book.....K}. 
Here we replace these with $m_0$ being the mass of the dark core and $r_0$ the Bondi radius, and we keep $l_0=0$. 
We make the code for computing these models, as well as for all the subsequent analysis, publicly available\footnote{\url{https://github.com/earlbellinger/darkcore} \citep{darkcore-zenodo}}. 

We calibrate each solar model by iteratively changing its initial helium abundance $Y_0$, initial metallicity $Z$, and mixing length parameter $\alpha_{\textrm{MLT}}$ until the model reaches the solar luminosity and solar radius at the solar age to a tolerance of within $10^{-7}$ of the respective values. 
The density distributions of the calibrated models are shown in Figure~\ref{fig:mr}. 
More details about the solar calibration procedure are given in Appendix~\ref{sec:calibration2}, and a visualization of the results are shown in Figure~\ref{fig:calibration}. 
\mb{For simplicity of the analysis, we assume a fiducial sound speed corresponding to the central value of the initial model and keep the corresponding Bondi radius fixed throughout the evolution.} 
\mbb{We also experimented with models having a variable Bondi radius, and found that its value changes by at most about an order of magnitude throughout the evolution due to the varying central sound speed. However, it was significantly more numerically challenging to converge these models, and therefore we restrict our attention in this work to models with a fixed inner boundary.} 
We furthermore tested the effect of varying the inner radius boundary condition, and therefore probing different dark matter densities, and found this had a very similar effect on the resulting solar structure as when changing the dark core mass. 

\begin{figure*}
    \centering
    \includegraphics[trim={0 0 68 0},clip,height=9cm]{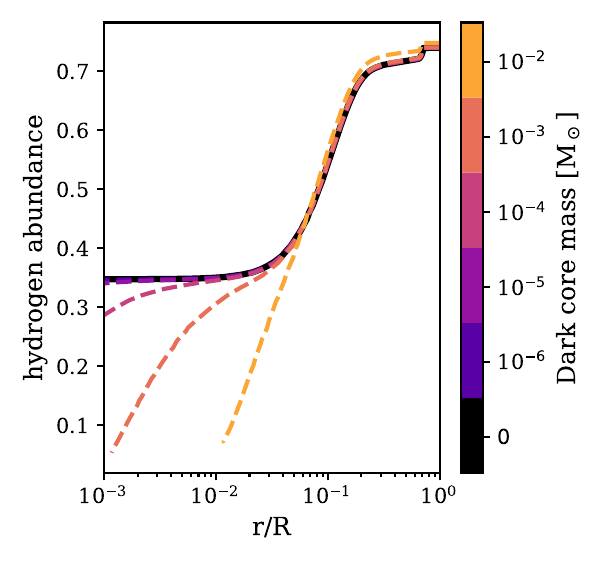}%
    \includegraphics[height=9cm]{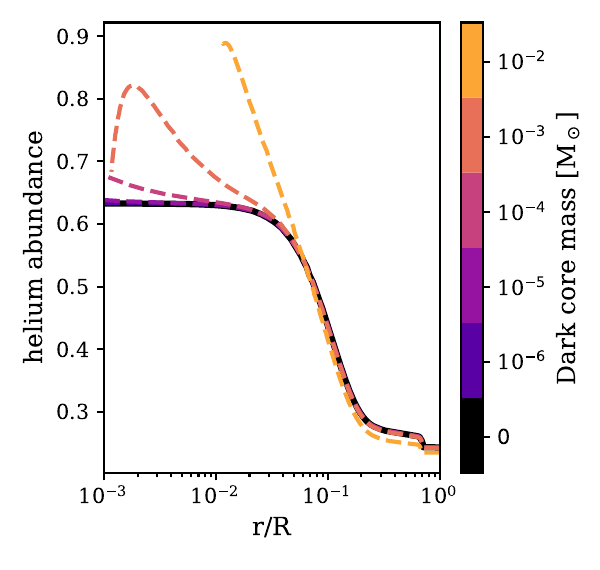}
    \caption{\textbf{Chemical abundances of dark core solar models.} 
    \textsc{Left Panel}: Hydrogen abundance profile. 
    \textsc{Right Panel}: Helium abundance profile. The increased central density leads to much more efficient central CNO burning in the massive dark core models, causing these models to be enhanced in central helium. Additionally, the strong gravitational potential transports metals to the deep core, causing near central hydrogen depletion at the solar age. \label{fig:abundances}}
\end{figure*}

The internal composition profiles of hydrogen and helium are shown in Figure~\ref{fig:abundances}. 
The chemical profile of the $10^{-3}~\rm{M}_\odot$ model is particularly noteworthy: its He abundance first increases and then decreases outward from the center. 
This arises due to dramatically enhanced settling of metals toward the solar center due to the much stronger gravitational potential caused by the dense dark core (see Appendix~\ref{sec:properties} for more visualizations). 
The innermost 0.1\% of this model by radius is more than $25\%$ metal, comprised mostly of $^{24}$Mg but also containing a substantial amount of $^{16}$O and $^{20}$Ne.

This mixture merits some interpretation, for which we will draw on the theory of settling in the interior of a white dwarf (WD). 
The settling force on a species $i$ is \citep[e.g.,][]{2020ApJ...902...93B}:
\begin{align}
    F_g 
    &= 
    -
    A_i m_p g 
    + 
    Z_i e E 
    \\
    &= 
    - 
    \left(
        A_i 
        - 
        Z_i 
        \frac{\langle A \rangle}
        {\langle Z \rangle}
    \right) 
    m_p g.
\end{align}
Here $A_i$ and $Z_i$ are the nuclear mass and charge numbers, with $m_p$ the nucleon mass, $g$ the inward gravitational acceleration, and $eE$ the outward electrostatic force for hydrostatic equilibrium. 
In a WD, ${\langle A \rangle / \langle Z \rangle\sim2}$, so typically only the neutron excess matters, causing nuclei like $^{22}$Ne and $^{25}$Mg to settle. In our simulations, ${\langle A \rangle / \langle Z \rangle}$ is biased upwards by large amounts of hydrogen, and so two effects emerge. 
The first is that symmetric $\alpha$-cluster nuclei sink, like $^{20}$Ne and $^{24}$Mg, and the second is that larger symmetric nuclei experience greater settling forces. 
It is worth noting that while $g$ goes effectively to zero in the limit of the $r=0$ in a WD, it does not with a dark core. Furthermore, $\langle A \rangle / \langle Z \rangle$ has a very strong time dependence due to the settling.

\section{Neutrinos} \label{sec:neutrino}
The central densities of the models scale strongly with the dark core mass, which profoundly impact the energy source of the star. Energy from the pp chain wanes in the center with increasing dark core mass, yet increases away from the center (see Appendix~\ref{sec:properties} for more visualizations). 
CNO burning becomes much more efficient in the central regions, causing a helium-rich core to form in models with a dark core mass exceeding $10^{-4}~\rm{M}_\odot$. 

\begin{figure*}
    \centering
    \includegraphics[trim={0 0 67 0},clip,height=9cm]{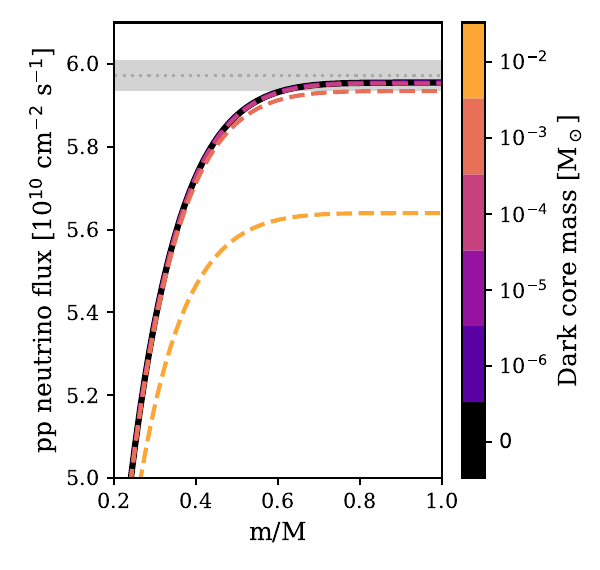}%
    \includegraphics[height=9cm]{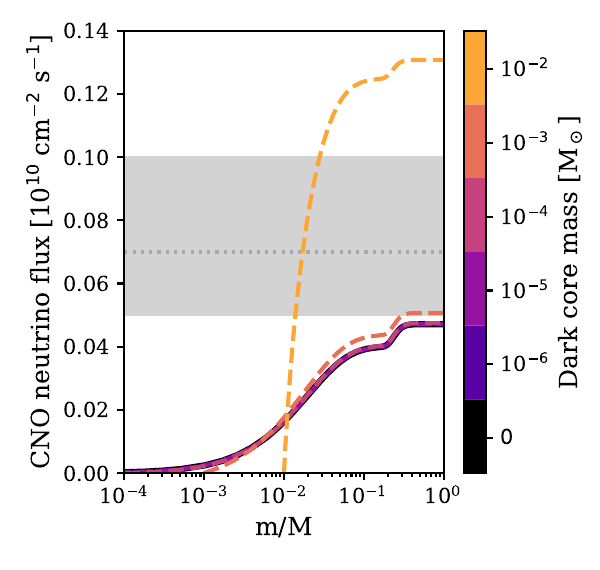}
    \caption{\textbf{Neutrino constraints on dark core masses.} 
    \textsc{Left Panel}: Neutrino flux from the fundamental reaction of the pp chain. 
    \textsc{Right Panel}: Net neutrino flux from the CNO cycle. The gray shaded regions give $1\sigma$ uncertainties from \citealt{2016JHEP...03..132B} and \citealt{2023PhRvD.108j2005B}, respectively. 
    In both cases, only the most massive dark core considered ($10^{-2}~\rm{M}_\odot$) can be ruled out using neutrino data. 
    \label{fig:neutrino}
    }
\end{figure*}

Figure~\ref{fig:neutrino} shows neutrino fluxes resulting from the increased central density. 
There is only significant change to the fluxes when the dark core mass is $\sim 1\%$ of the total solar mass. 
This makes sense when considering that the models are calibrated to produce $1~\rm{L}_\odot$ from nuclear reactions. 

\section{Helioseismology} \label{sec:helioseismology}
Here we examine the seismic properties of the dark core models. 
We use the open-source stellar oscillation code \textsc{Gyre} to compute the pulsation frequencies of the models \citep{Townsend+2013, Townsend+2018, Goldstein+2020, Sun+2023}. \mb{The stellar oscillation equations are normally solved with the inner boundary condition that $\xi_r-\ell\xi_h=0$, with $\xi_r$ the radial displacement, $\xi_h$ the horizontal displacement, and $\ell$ the spherical degree. 
However, this is only valid when $r_0=0$, and we have a dark core at the origin. 
We have thus searched for $p$ modes by enforcing $\xi_r=0$ at the dark core boundary, and $g$ modes with $\xi_h=0$.
We have moreover inspected the resulting eigenfunctions and eigenfrequencies related to assuming the opposite, and found that the conclusions of our subsequent analysis are the same with either choice. \mbb{As an example, the dipole $p$-mode frequencies of the least massive dark core model change by only $10^{-11}~\mu$Hz when switching between these assumptions, which is many orders of magnitude below observational uncertainty.}} 

\begin{figure*}
    \centering
    \includegraphics[trim={0 0 67 0},clip,height=9cm]{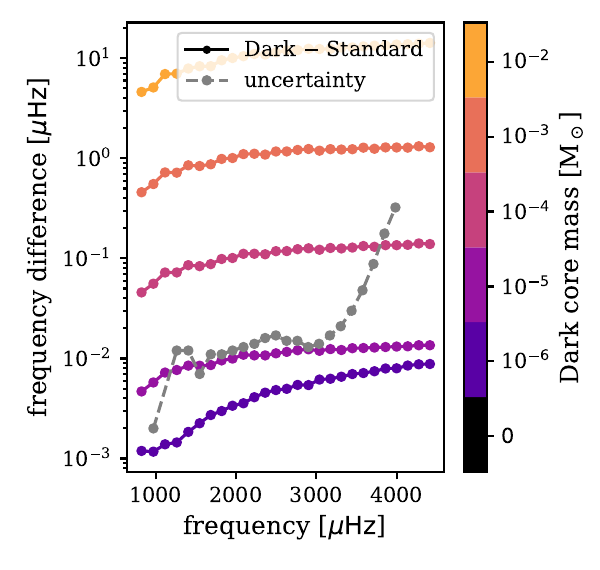}%
    \includegraphics[height=9cm]{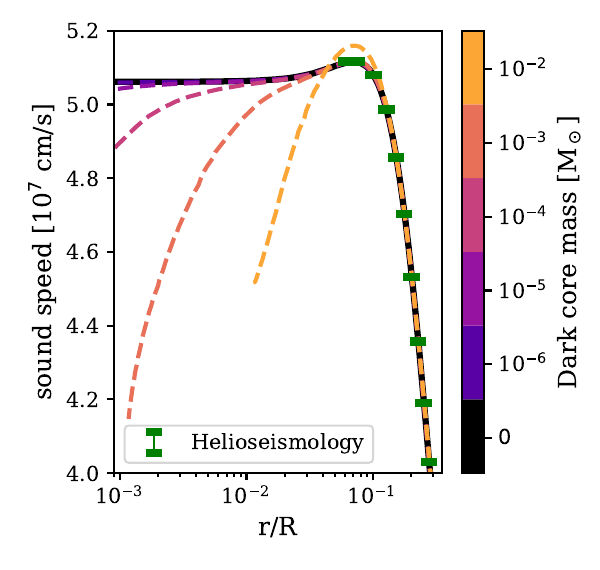}
    \caption{\textbf{Constraints from p-mode helioseismology on dark core masses.} 
    \textsc{Left Panel}: Theoretical differences in radial oscillation frequencies between a model with and without a dark core. The gray points give the 1$\sigma$ observational uncertainty of helioseismic data and thus show the potential to constrain models with masses greater than ${10^{-5}~\rm{M}_\odot}$. 
    \textsc{Right Panel}: Sound speed profiles of dark core solar models in the innermost regions. Helioseismic measurements of the solar sound speed profile from \citealt{2009ApJ...699.1403B} are shown in green. 
    \label{fig:csound}
    }
\end{figure*}

The Sun is observed to oscillate in a rich spectrum of acoustic modes, which reflect its internal sound speed profile \citep[e.g.,][]{2016LRSP...13....2B, 2021LRSP...18....2C}. 
Figure~\ref{fig:csound} shows the sound speed profiles of these models and their associated $p$-mode oscillation frequency differences, and a comparison with the uncertainties in the measured frequency as obtained by the BiSON solar observatory \citep{2014MNRAS.439.2025D}. 
Here it can be seen that current seismic measurements are capable of constraining the mass of a compact core to masses above $\sim 10^{-5}~\rm{M}_\odot$, three orders of magnitude stronger than neutrino flux measurements. 
Dark cores less massive than this limit are therefore not probed by $p$-mode helioseismology, and a more massive dark core than this limit could potentially be measured if one is indeed inside the Sun. 

\begin{figure*}
    \centering
    \includegraphics[trim={0 0 68 0},clip,height=9cm]{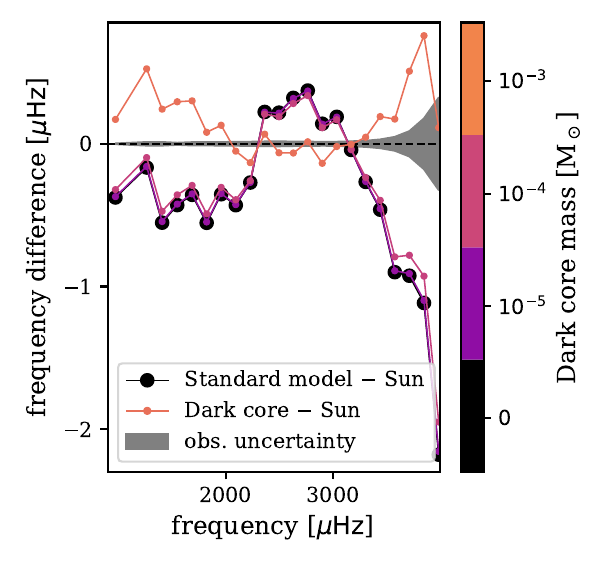}%
    \includegraphics[height=9cm]{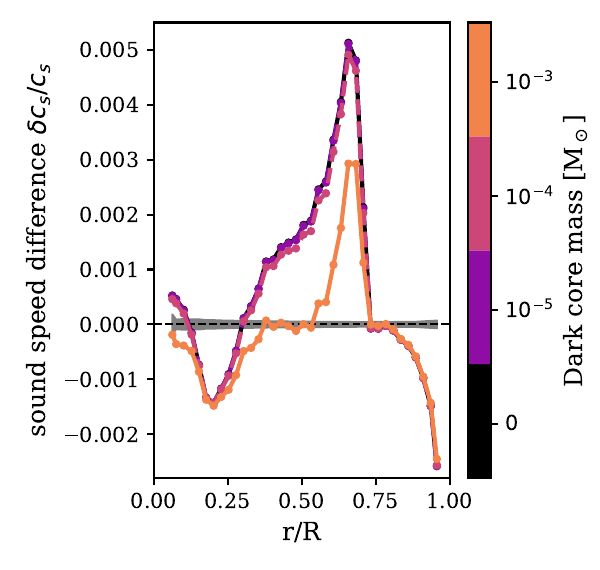}\\%
    \includegraphics[trim={0 0 68 0},clip,height=9cm]{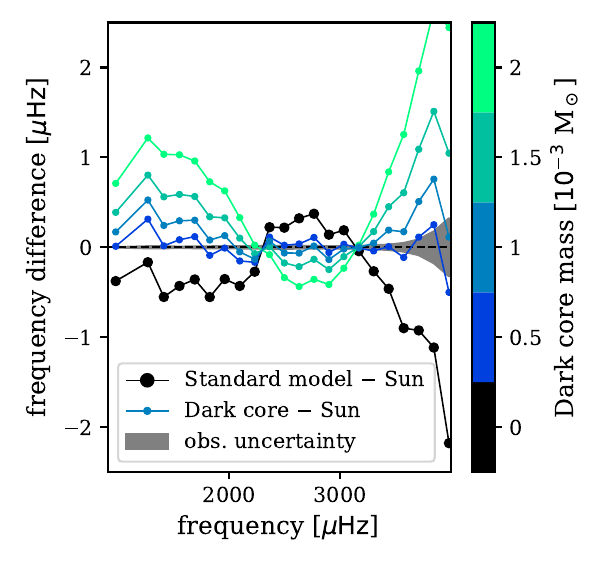}%
    \includegraphics[height=9cm]{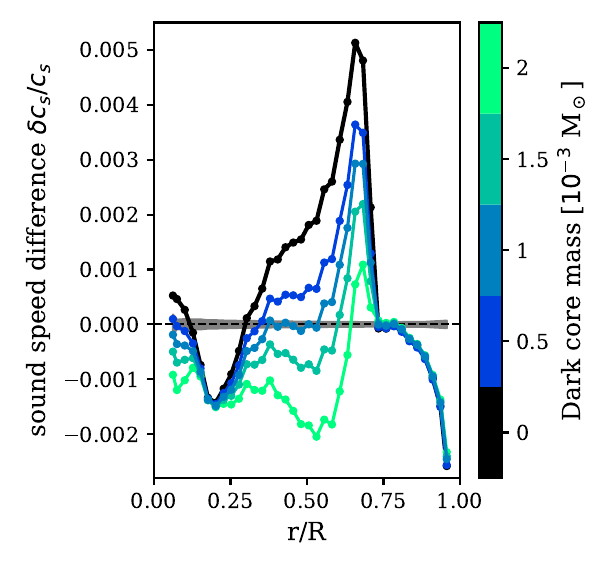}\\%
    \caption{\textbf{Comparison of dark core models with helioseismic observations.} 
    \textsc{Left Panels}: Comparison of dark core radial oscillation modes with helioseismic measurements, after correction for surface effects. The standard model of the Sun is discrepant with observations at a level of around $0.5~\mu$Hz. 
    Typical uncertainty is 0.01 $\mu$Hz. 
    \textsc{Right Panels}: Differences in internal sound speed profile between the models and the actual Sun, given in the sense of (Sun-model)/Sun. 
    The uncertainties of the measurements are around 0.005\% and hence are too small to be visible at the given resolution. 
    The standard solar model is significantly discrepant with the actual solar sound speed at a level of around 0.5\%. 
    The dark core model with $10^{-3}~\rm{M}_\odot$ appears to be in better agreement with observations. 
    The upper panels show a logarithmic spacing in dark core masses, while the lower panels show a linear spacing around the $10^{-3}~\rm{M}_\odot$ models.
    \label{fig:radial_sun}
    }
\end{figure*}

Now that we have assessed the theoretical precision of solar oscillations as a dark matter detector, we compare these dark core models with actual observations of the Sun. 
In Figure~\ref{fig:radial_sun} we compare the radial mode frequencies of the dark core models with the observed solar frequencies. 
In order to make this comparison, we first correct the radial mode frequencies for the surface effect using the treatment given by \citet{2014A&A...568A.123B}. 
In this Figure it is apparent that the dark core model with $10^{-3}~\rm{M}_\odot$ seems to improve the standard solar model. 

In order to assess the level of improvement, we compute $\chi^2 = \sum (\rm{model}-\rm{data})^2/\sigma^2$ and the reduced $\chi^2_{\rm{r}} = \chi^2/(\rm{\#~observations} - {\#~parameters})$ for the surface term corrected radial oscillation modes, where $\sigma$ is the uncertainty of each mode frequency. 
The standard solar model has $\chi^2_r=2633$ while the dark core model is around an order of magnitude better at $\chi^2_r=308$. 
We inspected modes of other degrees as well and similarly found general improvement. 
However, as will be explored in more detail in the next subsection, we also find mixed modes in the observable region. 
Depending on the mass of the dark core, mode mixing can cause the frequency of some non-radial modes to have even greater tension with the observations. 

In the right panel of Figure~\ref{fig:radial_sun} we furthermore compare the radial structure dark core models with that of the Sun. These measurements of the solar interior were inferred by \citet{2009ApJ...699.1403B} using linear perturbation theory applied to standard solar models. 
Here it can again be seen that the presence of a $10^{-3}~\rm{M}_\odot$ dark core appears to significantly improve the model's agreement with observations. 

Also observationally well-constrained quantities are the location of the base of the solar convection zone $x_{\rm{cz}} = {0.7133 \pm 0.0005}$ \citep{1998MNRAS.298..719B}, and the abundance of helium within the convection zone $Y_{\rm{cz}} = {0.249 \pm 0.003}$ \citep{1995MNRAS.276.1402B}. 
All the dark core models, including the standard solar model, have the nearly same value of $Y_{\rm{cz}}$ with only a slight decrease with increasing dark core mass. Except the most massive model, all the models are at within around 2$\sigma$ of the observed value. 
On the other hand, the standard model's value for $x_{\rm{cz}}$ is significantly too high by about $7\sigma$, while the value for the $10^{-3}~\rm{M}_\odot$ dark core is within $1\sigma$ of the observed value. 
Hence here again the dark core model appears to be in significantly better agreement with the helioseismic data.

These results are highly suggestive and merit some interpretation. 
We will begin by analyzing what about the structure of the dark core model has changed in order to improve the agreement with observations. 
The $p$ mode oscillations are sensitive to the speed of sound of the solar plasma, which in turn depends on the ratio of the temperature to the mean molecular weight $\mu$. 
The central temperature of this model is only slightly increased over the standard model, while $\mu$ is significantly enhanced in the innermost $1\%$ by radius due to the rapid settling of metals around the dark core (see Appendix~\ref{sec:properties} for more visualizations). 
Additionally, the size of the $p$ mode cavity is reduced by the radius of the Bondi sphere. 
In this model this is $0.001~\rm{R}_\odot$, which is about two orders of magnitude larger than the uncertainty on the solar radius \citep{2008ApJ...675L..53H}. 
Hence the combination of a heavy metal core and a reduced radius appears to drive the improvement. 

Recent research has indicated that the Sun may have formed with a metal-rich core and only later accreted metal-poor outer layers \citep{2019ApJ...881..103Z, 2021A&A...655A..51K}. 
If that hypothesis is correct, then the dark matter prescription in our models may be improving the agreement with the standard solar model by causing a similar effect. 
We now assess some effects that may rule out this model, or potentially constrain dark core masses even further.

\subsection{Mixed Modes} \label{sec:mixed-modes}
We find that in addition to $p$ and $g$ modes, dark core models oscillate in mixed modes that behave like a $p$ mode in the envelope of the star and like a $g$ mode in the deep interior. 
These modes are usually only observed in subgiants and red giants, which in those stars principally arise due to the increased buoyancy frequency in the hydrogen burning shell that emerges after the main sequence. 
In order to explain this behavior in dark core solar models, we analyze a \mb{differential equation that approximates the behavior} obeyed by stellar oscillations \citep[e.g.,][]{2010aste.book.....A}: 
    \begin{equation}
    \dfrac{\textrm{d}^2 \xi_r}{\textrm{d}r^2}
    =
    -K \xi_r
\end{equation}
where 
\begin{equation} \label{eq:frequencies}
    K
    =
    \dfrac{\omega^2}{c_s^2}
    \left(
        \dfrac{N^2}{\omega^2} - 1
    \right)
    \left(
        \dfrac{S_\ell^2}{\omega^2} - 1
    \right)
\end{equation}
with $\omega$ the oscillation frequency, $\xi_r$ the radial displacement, $r$ the distance from the stellar center, and $S^2_\ell = \ell\,(\ell+1)\,c_s^2 / r^2$, where $\ell$ is the spherical degree of the mode. 
Assuming an ideal gas, the buoyancy frequency is given by 
\begin{equation} \label{eq:N2}
    N^2 \approx \frac{g}{H_p} \left( \nabla_{\text{ad}} - \nabla + \nabla_\mu \right)
\end{equation}
with $g$ the gravitational acceleration, $\nabla_{\rm{ad}}$ the adiabatic temperature gradient, $\nabla$ the temperature gradient, and $\nabla_\mu$ the mean molecular weight gradient. 
The equations therefore permit two solutions: $p$~modes, where $\omega > S_\ell$ and $\omega > N$; and $g$~modes, where $\omega < S_\ell$ and $\omega < N$. A mode that satisfies the $g$~mode criterion in the interior layers and the $p$~mode criterion in exterior layers then oscillates as a mixed mode. 

\begin{figure*}
    \centering
    \includegraphics[trim={0 0 67 0},clip,height=9cm]{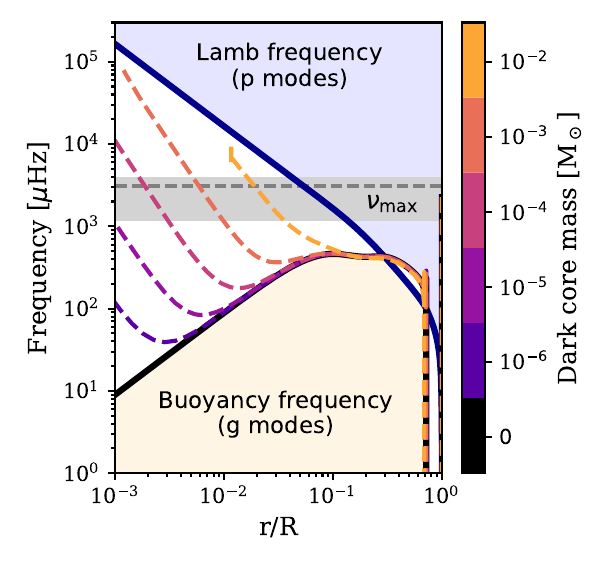}%
    \includegraphics[height=9cm]{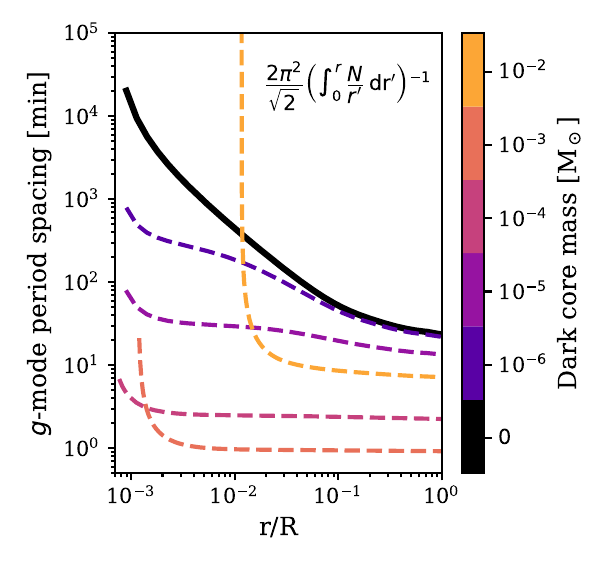}
    \caption{\textbf{Oscillation modes in dark core models.} \textsc{Left Panel}: Propagation diagram showing the \mb{dipole $p$-mode and $g$-mode} frequency cavities. Dark cores more massive than $10^{-5}~\rm{M}_\odot$ have an increased central buoyancy frequency, changing the $g$-mode period spacing and giving rise to observable mixed oscillation modes. 
    The grey shaded region gives range of observed solar oscillation frequencies. 
    For visualization purposes only the Lamb frequency profile of the standard model is shown as it is similar to the dark core models. 
    \textsc{Right Panel}: Asymptotic period spacings of \mb{dipole} $g$~modes in dark core models. The period spacings decrease with increasing dark core mass until the most massive, which has a higher period spacing due to its smaller cavity. 
    \label{fig:propagation}}
\end{figure*}

Figure~\ref{fig:propagation} shows a propagation diagram of the models, visualizing $S_1$ and $N$. 
The buoyancy frequency rises in the interior of the star due to the enhanced mean molecular weight gradient in the cusp (see Equation~\ref{eq:N2} and the chemical abundance profiles in Figure~\ref{fig:abundances}). 
This gives rise to mixed oscillation modes in the observable spectrum (indicated in the diagram by the frequency at maximum oscillation power $\nu_{\max}$ and the range of observed $p$~modes) for models with dark cores more massive than $10^{-5}~\rm{M}_\odot$. 

\begin{figure}
    \centering
    \includegraphics[width=\linewidth]{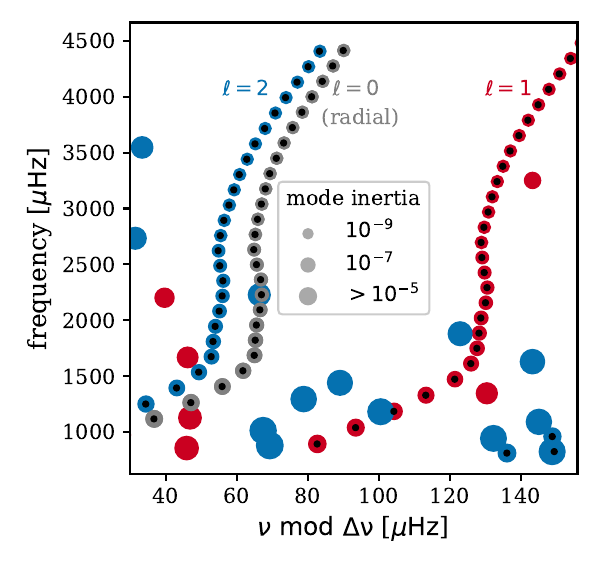}
    \caption{\'{E}chelle diagram comparing the $p$-mode frequency spectrum for a normal solar model (black points) and a model with a $10^{-4}~\rm{M}_\odot$ dark core (colored points). Mixed modes with very high mode inertia arise in models with dark core masses exceeding $10^{-5}~\rm{M}_\odot$. 
    Here we visualized the data using a large frequency separation of $\Delta\nu = 135.1~\mu$Hz corresponding to the observed solar value. \label{fig:echelle}}
\end{figure}

To explore this behavior more, we show an \mbb{\'{e}chelle} diagram in Figure~\ref{fig:echelle}. Also indicated here is the mode inertias, given by \citep[e.g.,][]{2010aste.book.....A}%
\begin{equation}
    E
    =
    \frac{4\pi\int_0^R\left[|\xi_r|^2+\ell(\ell+1)|\xi_h|^2\right]\rho
    r^2\, \textrm{d}r}
    {\left[|\xi_r(\textrm{R}_\odot)|^2+\ell(\ell+1)|\xi_h(\textrm{R}_\odot)|^2\right] \textrm{M}_\odot}\text{,}
\end{equation}
with $\xi_r$ and $\xi_h$ being the radial and horizontal displacement eigenfunctions for that particular oscillation mode. 
The dark core mixed modes generally have inerti\ae{} that are orders of magnitude larger than the $p$~modes. 
Modes with high inertia are trapped deeper within the interior of the star and are hence thought to be less observable at the surface. 
However, computation of stellar oscillation amplitudes in solar-like stars is an unsolved problem, and therefore we are unable to assess whether these modes would be presently detectable. 
\mbb{Similarly, although mode visibilities in the geometric sense can be computed, quantitative predictions of surface amplitudes---especially for mixed or $g$ modes---remain uncertain. As such, we cannot assess their detectability from visibility arguments alone.} 
If these modes are indeed detectable, their lack of detection in solar observations constrains the dark core mass to at most $10^{-5}~\rm{M}_\odot$. 

Upon general inspection, the high-inertia mixed modes in these models are not obviously incompatible with the observations. 
However, it should be noted that while most of the mixed modes appear as extra modes (as in Figure~\ref{fig:echelle}), some mixed modes in the $10^{-3}$~M$_\odot$ dark core model take the place of the ordinary $p$ modes in the observable region, thereby placing them in high tension with the observations. 
Thus despite the other apparent improvements to the model, mixed modes may rule out such a massive dark core. 
It may be possible to optimize the dark core mass to find a model in even better agreement with the observations, but we elect not to do this here. 
That being said, the general lack of mixed modes observed in the Sun and main-sequence solar-like oscillators alike likely indicates that dark cores of this mass are absent or rare.

\section{Solar g~modes} \label{sec:g-modes}
We finally turn our attention to the probing power of solar $g$~modes. 
Despite decades of search, these modes have so far eluded detection due to their extremely small amplitudes, which are estimated to be at most $1~\rm{cm}/\rm{s}$ \citep[e.g.,][]{2010A&ARv..18..197A}. 
Nevertheless, unlike $p$~modes, which reveal down to the innermost few percent of the solar structure, discovery of solar $g$~modes promise to reveal the detailed structure around the stellar center. 
In the limit of high radial order, the periods of $g$~modes are spaced to leading order according to 
\begin{equation} \label{eq:per-spac}
    \Delta \Pi_\ell \simeq \dfrac{2\pi^2}{\sqrt{\ell\,(\ell+1)}}
    \left(
        \int_{N^2 > 0} \dfrac{N}{r} \; \textrm{d}r
    \right)^{-1}.
\end{equation}
Shown in the right panel of Figure~\ref{fig:propagation} is the cumulative integral of $\Delta\Pi$ for $\ell=1$ (dipole) $g$~modes, with the actual spacing given by the value at the surface. 
Here an interesting non-monotonicity can be observed. 
The asymptotic period spacing first decreases with increasing dark core mass from the normal $25$~min down to a minimum of $1$~min, caused by the increasing values of $\nabla_\mu$. It then increases again for the $10^{-2}~\rm{M}_\odot$ model, which is caused by the shrinking radiative zone, as modes in this model have a propagation cavity that is more than $1\%$ smaller than the others. 

\begin{figure*}
    \centering
    \includegraphics[trim={0 0 67 0},clip,height=9cm]{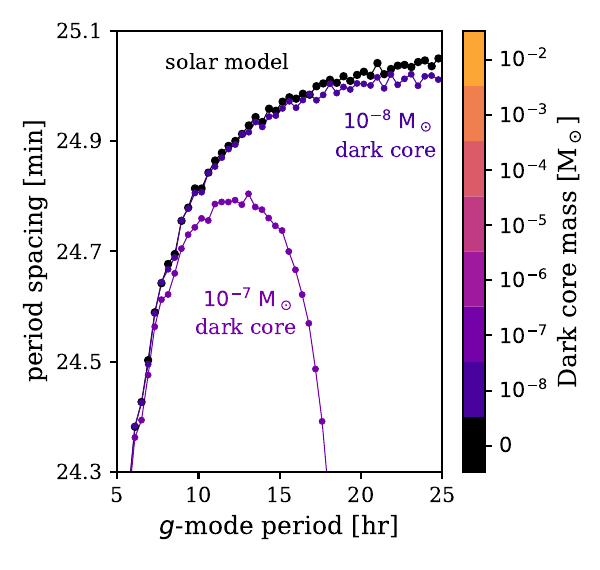}%
    \includegraphics[height=9cm]{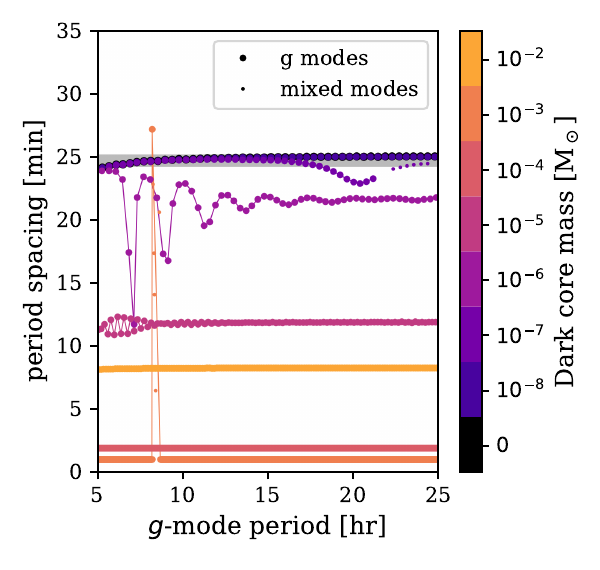} 
    \caption{\textbf{Helioseismology of dark core $g$~modes.} 
    \textsc{Left Panel}: The \mb{dipole} $g$-mode period spacings of the normal solar model as well as the effects of the two least massive dark core models. If $g$~modes can be observed, they may constrain dark core masses down to $10^{-7}~\rm{M}_\odot$. 
    \textsc{Right Panel}: \mb{Dipole} period spacings of all the models. The gray background depicts the area shown in the left panel. Models with massive dark cores ($\geq 10^{-4}~\rm{M}_\odot$) have regularly spaced $g$~modes with a small period spacing. Intermediate-mass dark cores ($10^{-5}-10^{-6}~\rm{M}_\odot$) are dominated by mixed oscillation modes, disturbing any regular spacing. 
    \label{fig:gmodes}
    }
\end{figure*}

Lastly, Figure~\ref{fig:gmodes} shows the periods and period spacings of the $g$~modes in the dark core models. 
The non-monotonicity previously observed in the asymptotic period spacing is borne out in the calculation of the actual mode periods. 
\mbb{Some of the models show oscillations in their period spacings, which likely arise due to the sensitivity of the buoyancy frequency to the mean molecular weight gradient (e.g., \citealt{2008MNRAS.386.1487M}; see also Figure~\ref{fig:metals}). In addition, the models exhibit modes labeled as mixed modes at long period, which likely arise due to self-interactions of $g$ modes as observed in models of more massive stars \citep[e.g.,][]{2023A&A...675A..17V}.} 

Here we also find the strongest constraint. 
The spacings of long-period $g$~modes are significantly perturbed at the level of $2\%$ by a dark core of $10^{-7}~\rm{M}_\odot$. 
If the precision of solar $g$~modes, when they are finally observed, is comparable to the precision with which solar $p$~modes are measured, then this would be a measurable effect. 
That being said, it is conceivable that other uncertainties in the solar physics assumed in the simulation, such as the treatment of the shear layer at the base of the convection zone, may lead to theoretical $g$~mode uncertainties that vary by larger amounts, but we have not assessed any such effects.

\section{Conclusions} \label{sec:conclusions}
We have quantified how a compact dark object in the center of the Sun would alter its observable properties. 
Solar neutrino measurements rule out dark core masses at or above $10^{-2}~\rm{M}_\odot$ (see Figure~\ref{fig:neutrino}). 
Solar $p$~mode frequency measurements currently constrain dark cores at a precision of $10^{-5}~\rm{M}_\odot$ (see Figure~\ref{fig:csound}). 

We have found that a model with a $10^{-3}~\rm{M}_\odot$ dark core is in better agreement with helioseismic measurements than the standard solar model (see Figure~\ref{fig:radial_sun}). 
We suggest this arises because the dark matter models have heavy metal cores, induced by the more efficient settling caused by the enhanced gravitational potential in the central regions of the model. 
Recently, \citet{2019ApJ...881..103Z} and \citet{2021A&A...655A..51K} have suggested that the Sun may have formed with a metal-rich core. 
We therefore find it plausible that the dark core prescription could be improving the model because it is emulating these otherwise neglected effects of star formation. 
Also, we add that it is relatively unsurprising that models improve with the introduction of an additional free parameter, and though the amount of improvement caused by the dark core is itself perhaps surprising, the burden of evidence for finding dark matter in the Sun should be high. 
Accommodating the existing data is probably insufficient to be convincing; the model should also make additional predictions. 

It is therefore interesting to contemplate what effects could differentiate a dark core from the effects of star formation, or otherwise constrain its possible mass further. 
The dark core model additionally differs from the standard model in that it has mixed oscillation modes \mb{(see Figures~\ref{fig:propagation} and \ref{fig:echelle})}, which have not been observed in the solar data. 
However, these mixed modes generally appear as additional modes with high mode inertia, and therefore may be as difficult to observe as the so-far elusive solar $g$ modes. 
It is challenging to assess the amplitude of the modes, but if it is determined that these modes would be observable, they would rule out dark cores above $10^{-5}~\rm{M}_\odot$. 

At higher mass, some of the the observable non-radial $p$ modes also become mixed, which may rule out the $10^{-3}~\rm{M}_\odot$ model which otherwise seems to improve the solar model. 
It may be possible to tune the dark core mass such that this behavior is avoided for the solar model while still generally improving it, but the non-observation of mixed modes in other solar-like oscillators in general may indicate that such dark cores are non-existent or rare. 
On the other hand, the sample of known solar-like oscillators is small 
\citep[$\sim$100, e.g.][]{2017ApJ...835..172L, 2019A&A...622A.130B}. 
\mb{Much greater is the number of known stars with mixed modes}, which appear in the oscillation spectrum at the end of the main sequence once shell fusion becomes efficient \citep[e.g.,][]{1977A&A....58...41A, 2019MNRAS.489..928S, 2021ApJ...915..100B}. 
The sample of solar-like oscillators stands to be greatly expanded with the forthcoming PLATO mission \citep{2024arXiv240605447R} which may present further opportunities to study dark cores. 

Lastly, we have assessed the effects of solar $g$~mode measurements, should they one day come. 
These modes are roughly evenly spaced by about $25$~min in the standard solar model (see Figure~\ref{fig:gmodes}). 
This pattern is severely perturbed by masses as small as $10^{-7}~\rm{M}_\odot$. 
If the $10^{-3}~\rm{M}_\odot$ dark core model were in fact the situation for our Sun, its periods would be spaced significantly more closely, at a separation of about $1$~min. 
Such a prediction would confirm or rule out a dark core, and also distinguish it from other scenarios, such as a heavy metal core arising from star formation. 

Determining exactly which dark matter candidates these investigations constrain or support is beyond the scope of this article.
A few remarks are nevertheless possible. 
\citet{2020JCAP...09..044B} studied dark massive compact halo objects (dMACHOs) and found that they could span the mass range $10^{-12}-10$~M$_\odot$ and comprise all the dark matter. Hence solar observations may constrain this candidate. 
A separate candidate also belonging to the class of macroscopic dark matter is dark quark nuggets. \citet{2019PhRvD..99e5047B} studied these and found them to be constrained by microlensing to a much more limited mass range of up to perhaps $10^{-10}$~M$_\odot$, which is less massive than is even probed by solar $g$ modes. 
Hence unless the capture rate is large, or their mass distribution has a massive tail, it is unlikely that solar data can offer further constraints to that candidate. 

Primordial black holes, especially in the asteroid-mass window, remain a popular dark matter candidate. 
At small mass, the conditions of the simulations analyzed here are approximately appropriate for a primordial black hole \citep[PBH,][]{1971MNRAS.152...75H, 2024PhR..1054....1C} at the center of the Sun, and may even be appropriate for a larger-mass PBH if the accretion onto the hole is radiatively inefficient \citep{1995MNRAS.277...25M, bellinger2023solar, caplan2024there, 2024arXiv240617052S}, \mb{i.e., if most photons are not able to escape the Bondi radius and thus the accretion mainly results in the growth of the black hole}. 
The simulations considered here do not include mass accretion, meaning that the amount of gravitational settling would be overestimated for the PBH case. 
A fine tuning argument against a radiatively-efficient PBH in the solar core was given by \citet{{1995MNRAS.277...25M}}: if such a PBH in the mass range considered were indeed lurking in the Sun, the end would be near; and it is rather implausible that humanity would discover the tools to find a black hole in the solar center only moments before it ushers the Sun's demise. 
\cite{bellinger2023solar} considered the evolution of radiatively-efficient PBHs in stars, \mb{i.e., a case in which the accretion onto the black hole is luminous and thus serves as an energy source for the star. Such black holes consume the star much more slowly due to the feedback on the stellar structure. Their simulations constrained such a black hole mass inside the Sun to 10$^{-6}~\rm{M}_\odot$.} 
These ``Hawking stars'' transform into a rare but observed type of red giant called a red straggler, for which standard stellar evolution provides no recipe. 

A final consideration is that even if PBHs do supply all the dark matter, the expected stellar capture rate of PBHs in the Milky Way is very low, with a probability of only about 1 in 10 million that the Sun would have captured one \citep{2023PhRvD.107j3052E}. 
It is conceivable that stellar capture rates of other macroscopic dark matter candidates could be similarly low, and the event of their capture less dramatic. 
On the other hand, black holes are maximally compact and thus have very little cross-section for interaction, making their capture less likely. 
Other dark matter candidates of macroscopic mass and less compactness may therefore be more favorable for stellar capture. 
That said, if the mass of the DM candidate is large, then the overall number of them must be smaller, thus also reducing the chance of stellar capture. 
More generally, it is conceivable that some fraction of stars may have captured some amount of dark matter, and others a negligible amount or none at all. 
Into which category our own Sun falls will require further study and observation.

\begin{acknowledgments}
We are grateful to Sarbani Basu, J{\o}rgen Christensen-Dalsgaard, Ebraheem Farag, and the anonymous referee for their helpful feedback and discussions that improved the manuscript. 
\end{acknowledgments}

\appendix

\section{Solar calibration} \label{sec:calibration2}
\begin{figure*}
    \centering
    \includegraphics[trim={0 0 67 0},clip,height=9cm]{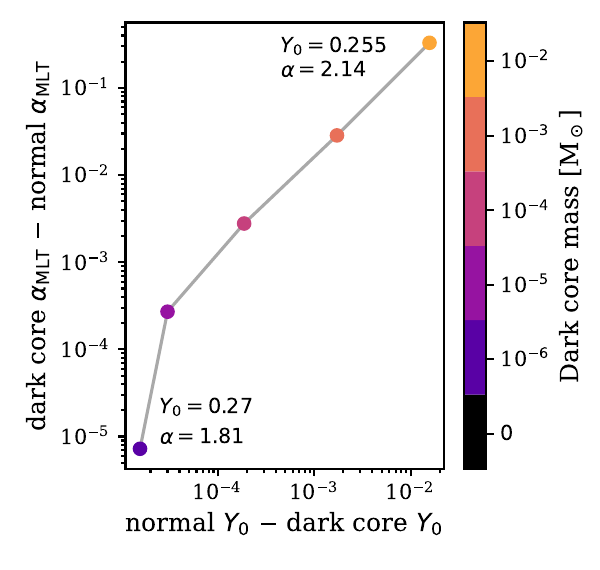}%
    \includegraphics[height=9cm]{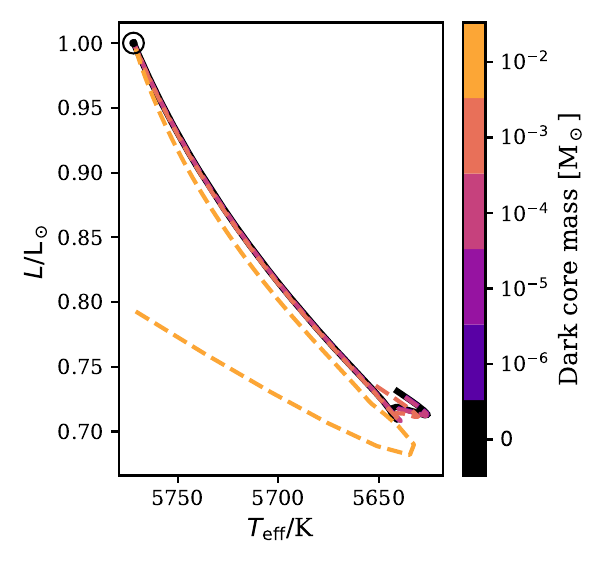}
    \caption{\textbf{Solar calibration of dark core models.} \textsc{Left Panel}: Initial parameters (initial helium abundance $Y_0$ and mixing length parameter $\alpha_{\rm{MLT}}$) of dark core models that yield models of the present Sun. 
    \mb{Calibration parameters for the least and most massive dark cores shown are labeled on the figure; the least massive dark core has $Y_0$ and $\alpha_{\rm{MLT}}$ values that are within $10^{-4}$ of those of standard model.} 
    Increasing the dark core mass requires a decrease to $Y_0$ and increase to $\alpha_{\rm{MLT}}$ to achieve the solar radius, luminosity, and metallicity at the solar age. \textsc{Right Panel}: Hertzsprung--Russell diagram, verifying that the models all reach the solar position. Each model is colored by its dark core mass as given in the colorbar. Note that the most massive dark core model approaches the zero-age main sequence from a higher effective temperature than the models with less massive cores, \mbb{which is caused because unlike the other masses, its initial model is already gaining a substantial fraction of its energy from pp and CNO fusion reactions.} \label{fig:calibration}}
\end{figure*}

Here we give a few more details on the solar calibration. 
We assume a fiducial solar radius of $6.9566\times10^{10}$ \citep{2008ApJ...675L..53H} and luminosity of $3.828\times10^{33}$, and a surface value of $(Z/X)=0.02293$ corresponding to the \citet{1998SSRv...85..161G} solar composition. 
We use $G = 6.674484 \times 10^{-8}$ in cgs units \citep{2018Natur.560..582L} and $\rm{M}_\odot = 1.988355 \times 10^{33}$~g \citep{2021AJ....161..105P}.
Figure~\ref{fig:calibration} shows a verification of the solar calibration process, and the resulting dark core input parameters. 

\section{Model Properties} \label{sec:properties}
\begin{figure*}
    \centering
    \includegraphics[trim={0 0 67 0},clip,height=9cm]{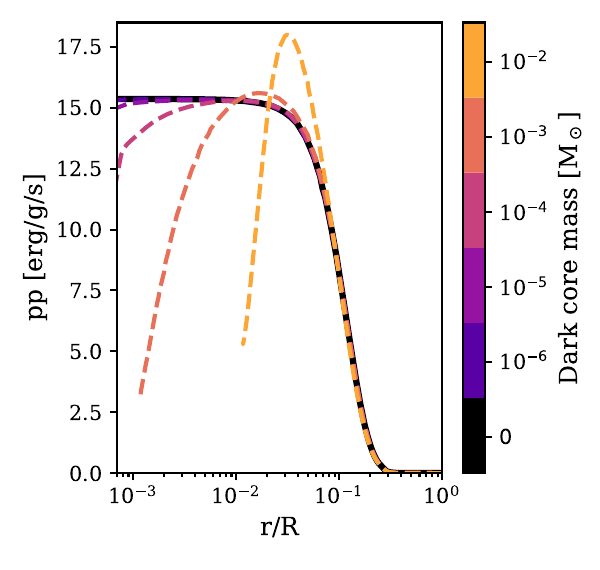}%
    \includegraphics[height=9cm]{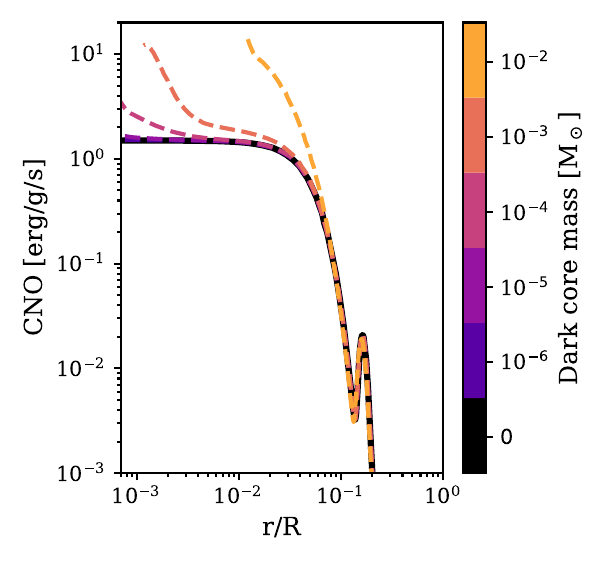}
    \caption{\textbf{Nuclear energy sources in dark core models.} \textsc{Left Panel}: Energy generation from the pp chain \mb{in the present Sun}. \textsc{Right Panel}: Energy generation from the CNO cycle.  \label{fig:nuc}}
\end{figure*}

\begin{figure*}
    \centering
    \includegraphics[trim={0 0 67 0},clip,height=9cm]{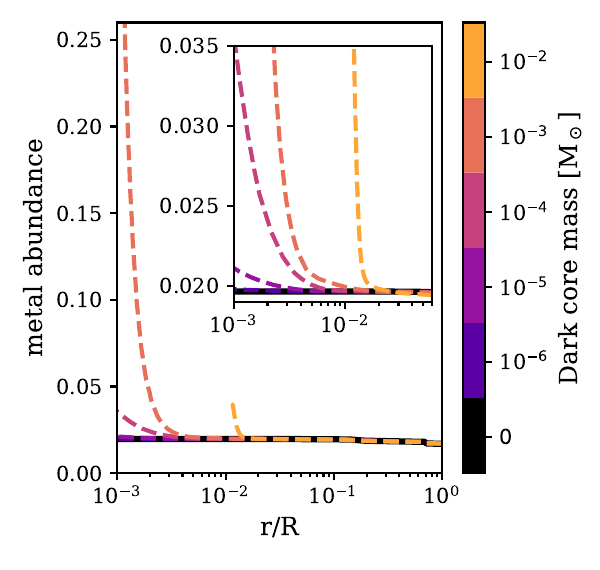}%
    \includegraphics[height=9cm]{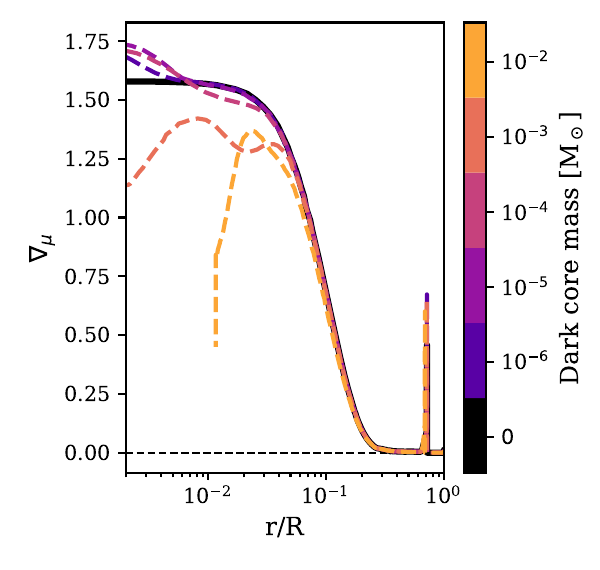}
    \caption{\textbf{Dark core heavy metal.} \textsc{Left Panel}: Mass fractions of metals in dark core models. \textsc{Right Panel}: Mean molecular weight gradient in dark core models. \label{fig:metals}}
\end{figure*}

Here we show a few auxiliary plots that help illuminate some properties of the models. 
Figure~\ref{fig:nuc} shows the energy generation from the pp chain and CNO cycle throughout the interiors of the models. 
Figure~\ref{fig:metals} shows the metallicity profile and the mean molecular weight gradient of the models. 

\bibliography{sample631}{}
\bibliographystyle{aasjournal}

\end{document}